\documentclass[lettersize,journal]{IEEEtran}
\usepackage{amsmath,amsfonts}
\usepackage{algorithmic}
\usepackage{algorithm}
\usepackage{array}
\usepackage[caption=false,font=normalsize,labelfont=sf,textfont=sf]{subfig}
\usepackage{textcomp}
\usepackage{stfloats}
\usepackage{color}
\usepackage{url}
\usepackage{verbatim}
\usepackage{graphicx}
\usepackage{cite}
\begin{document}

\title{Sequence Selection with Dispersion-Aware Metric for Long-Haul Transmission Systems}

\author{Jingtian Liu\textsuperscript{(1)}, Élie Awwad\textsuperscript{(1)}, Hartmut Hafermann\textsuperscript{(2)}, Yves Jaou\"{e}n\textsuperscript{(1)}
\thanks{E-mail of corresponding author: jingtian.liu@telecom-paris.fr}
\thanks{\textsuperscript{(1)} Communication and Electronics department, LTCI, Télécom Paris, Institut Polytechnique de Paris, 19 place Marguerite Perey, 91120 Palaiseau, France;~~\textsuperscript{(2)} Optical Communication Technology Lab, Paris Research Center, Huawei Technologies, France.}}


\IEEEpubid{}

\maketitle

\begin{abstract}

We introduce a novel sign-dependent metric: the energy dispersion index (EDI) of sequences that endured chromatic dispersion, denoted as D-EDI, which exhibits a more accurate opposite variations with the transmission performance compared to the standard EDI metric. Then, by applying D-EDI and EDI to the sequence selection (SS) process, 
we present two signaling approaches denoted as D-SS and E-SS respectively. These approaches are designed to minimize rate loss and enhance transmission performance in nonlinear optical fiber transmission systems, catering to both short-distance and long-haul scenarios. With enumerative sphere shaping (ESS) as distribution matcher (DM), our simulation results reveal significant performance gains over ESS without sequence selection (SS), with improvements up to $0.4$~bits/4D-symbol. These improvements were observed over a $205$-km single-span standard single mode fiber link in WDM transmission, with five dual-polarization channels, each operating at a net rate of $400$~Gbit/s. Furthermore, we demonstrate that D-SS surpasses ESS without SS by $0.03$~bits/4D-symbol in achievable information rate over a $30\times80$~km link in a single-wavelength, with 8 discrete multi-band (DMB) transmission, and an $880$~Gbit/s net rate. Notably, our proposed D-SS scheme achieves similar performance to a sequence selection based on a full split-step Fourier method (SSFM) simulation and it consistently delivers throughput enhancements across various block lengths and selected sequence lengths.
\end{abstract}


\section{Introduction}

Numerous studies have shown that achieving a high throughput in optical fiber transmission systems is feasible through the utilization of polarization division multiplexing (PDM), combined with $M$-ary quadrature amplitude modulation (QAM) formats and forward error correction (FEC) codes~\cite{pfau2009hardware,Onohara10,Kikuchi16}. However, this increased efficiency often comes at the cost of reduced transmission distances due to OSNR limitation and Kerr-induced nonlinear distortions~\cite{Frey17}. To approach the Shannon limit over a channel with additive white Gaussian noise (AWGN), constellation shaping has emerged as a promising strategy. Kschischang and Pasupathy's work demonstrated that the probability distribution that maximizes the achievable information rate over an AWGN channel, while maintaining a fixed average energy constraint, corresponds to the Maxwell-Boltzmann (MB) distribution~\cite{kschischang1993optimal,buchali2015rate,fehenberger2016probabilistic}.

To effectively implement the MB distribution, various distribution matching (DM) techniques have been proposed for probabilistic constellation shaping (PCS). Notable methods include enumerative sphere shaping (ESS)~\cite{gultekin2019enumerative} and constant composition distribution matching (CCDM)~\cite{schulte2015constant}. CCDM~\cite{schulte2015constant} has emerged as a competitive technique for implementing PCS with $M$-QAM constellations by delivering near-optimal linear shaping gains when large amplitude blocks are considered~\cite{alvarado2017achievable}. However, in the context of mitigating Kerr-induced nonlinear interference (NLI) in optical fiber transmission, long-block-length DM may exhibit a higher susceptibility to performance degradation, as highlighted in previous analyses. In~\cite{fehenberger2019analysis,fehenberger2020mitigating,gultekin2022optimum}, it was shown that short-block-length DM can effectively reduce NLI. Yet, it is crucial to recall that, as the block length decreases, the rate loss increases. Hence, finding a high-performance nonlinear-tolerant shaping scheme involves striking a trade-off between nonlinear gain and rate loss~\cite{amari2019introducing}. In this respect, ESS is more favored compared to CCDM due to its lower rate loss at the same short block lengths. Furthermore, simulation results in~\cite{civelli2023nonlinear,borujeny2023constant} showed that carrier phase recovery (CPR) in the DSP chain at the receiver side can exhibit comparable effectiveness to short-length PCS in mitigating NLI.

To further enhance the nonlinear performance of PCS schemes, recent research efforts focused on modifying the DM structure. For ESS, the nonlinear tolerance of its output amplitude sequence was improved by introducing constraints in the grid tree of the ESS, such as the kurtosis-limited ESS (K-ESS) in~\cite{gultekin2021kurtosis} and the band-Limited ESS (B-ESS)~\cite{gultekin2022mitigating}. In both approaches, 1D amplitude variance and/or kurtosis were limited, wich offers a nonlinear gain. This effect is underpinned by the EGN model~\cite{carena2014egn} that forecasts a more significant nonlinear penalty associated with higher kurtosis levels. While both K-ESS and B-ESS implement sequences in four-dimensional (4D) real-valued space while applying energy limitations over each dimension separately, we proposed in~\cite{Liu23} a band-limited version of ESS with energy limitations in 4D (two pairs of I and Q signals), achieving higher throughput. However, for these three schemes, the rate loss was still significant.


Other design approaches of nonlinear tolerant signaling schemes consist of manipulating the input signal to the DM to generate multiple candidate sequences. Then, specific metrics are used to select ``good" sequences, i.e. sequences that generate low NLI levels. For example, the energy dispersion index (EDI), introduced in~\cite{wu2021temporal}, is a metric that quantifies energy changes of the signal over a given temporal window. EDI showed an opposite variation with the effective signal-to-noise ratio (SNR) of the CCDM-based transmission. However, this behavior was observed in a single-polarization transmission scenario. As for ESS-based schemes, our study in~\cite{Liu23} highlighted that the EDI, when calculated over 4D-energy levels, demonstrates a notable opposite variation with SNR for ESS, B-ESS~\cite{gultekin2022mitigating}, and 4D-BL-ESS over $110$~Gbaud, $400$~km, $5$~WDM transmission. However, this behavior appeared to be inaccurate in the case of K-ESS~\cite{gultekin2021kurtosis}. In a different work~\cite{wu2022list}, EDI was utilized for sequence selection in a scheme called list-encoding CCDM (L-CCDM), achieving nonlinear gains with respect to conventional  CCDM. Yet, for long-haul transmissions, CCDM is not the best distribution matcher, as similar nonlinear gains can be achieved through ESS with shorter block lengths~\cite{amari2019introducing}. More recently, yet another approach~\cite{askari2023probabilistic} used EDI and a perturbation-model-based low-pass-filtered symbol-amplitude sequence (LSAS) metric~\cite{askari2022nonlinearity} for sequence selection. These methods have substantially improved the nonlinear tolerance of ESS or CCDM. However, it is essential to highlight that the most significant gains in ESS schemes have been observed over short-distance single-span transmissions.

The concept of sequence selection (SS) was first introduced in~\cite{civelli2021sequence} and has been further studied in~\cite{secondini2022new}. This innovative signaling approach provided valuable insights into the computation of a lower bound of the optical fiber channel capacity. In particular, the analysis clearly showed that there is considerable room for improvement in optical fiber transmission, especially for long-haul links. A recent study ~\cite{civelli2023sequence} from the same authors emphasized the importance of performing sequence selection using sign-dependent metrics in long-haul transmission scenarios. The authors demonstrated that sign-independent metrics such as EDI, windowed kurtosis, or LSAS do not yield significant throughput improvements over long-haul links. They also showed that a sequence selection based on the computation of NLI  through a numerical simulation (split-step Fourier method or SSFM) of a noiseless single-channel propagation of sequences achieved throughput enhancements over long-haul transmissions even when an optimized CPR is applied. However, this method is not feasible in practice due to the high complexity of the SSFM.


In this paper, we introduce a novel sign-dependent metric, the EDI of dispersed sequences, named D-EDI, which accounts for the influence of chromatic dispersion throughout transmission. Additionally, we show that D-EDI demonstrates a negative correlation with the effective SNR for high-rate multi-span scenarios. By negative correlation, we mean that the two metrics vary in opposite directions. Next, we utilize EDI and D-EDI as metrics for sequence selection, with ESS serving as the DM. The approach that employs sign-independent EDI for sequence selection is termed E-SS, and the one using sign-dependent D-EDI is designated as D-SS\footnote{For a PCS system employing sequence selection based on either EDI or D-EDI and comprising $ n $ cascaded DMs, where each DM includes $ \nu $ flipping bits, we name the scheme $ \mathrm{E\text{-}SS_{n}^{\nu}} $ for the EDI-based approach, and $ \mathrm{D\text{-}SS_{n}^{\nu}} $ for the D-EDI-based approach. The schemes will be detailed in section~\ref{E-SS and D-SS design}.}. We thoroughly investigate and optimize the performance of these schemes in both single-span and multi-span transmission scenarios with optimized CPR. Our D-SS scheme demonstrates superior performance compared to ESS without sequence selection across various block lengths and complexity levels, in both single-channel and WDM transmission scenarios. Remarkably, it performs on par with the ideal SSFM-based sequence selection method, however with a reduced complexity. The paper is structured as follows: in Section~\ref{System_Description}, we describe the simulated transmission systems, then, in Section~\ref{EDI_TO_D-EDI}, we define the new metric D-EDI. In Section~\ref{E-SS and D-SS design}, we outline the architecture of the E-SS and D-SS transmitter schemes. Section~\ref{Single span transmission} focuses on the performance of a single-span transmission system and gives insights into the optimization of the proposed methods in this context. In Section~\ref{Multi span transmission}, we study the performance over multi-span single-channel and WDM transmission systems with CPR at the receiver side. Finally, Section~\ref{Conclusion} summarizes key findings and conclusions drawn from our investigations.

\section{System Description}\label{System_Description}

We study the performance of E-SS and D-SS in two scenarios: first, a single-span transmission scenario similar to the one in~\cite{gultekin2021kurtosis,gultekin2022mitigating}; second, a multi-span long-haul transmission using digital-multi-band (DMB) format. Through the two studies, we observe the effects of nonlinear distortions with various signaling schemes. NLI distortions are dominant at high power levels, therefore, they primarily occur over the first kilometers of the span in a single-span transmission, and occur at the start of each span in a multi-span scenario with discrete optical amplification.

Our simulated single-span scenario consists of a transmission over $205~\mathrm{km}$ of standard single-mode fiber (SSMF), to perform a consistent comparison with~\cite{gultekin2021kurtosis,gultekin2022mitigating} in which the kurtosis-limited (K-ESS) and band-limited (B-ESS) sphere shaping were introduced. We perform probabilistic shaping using PDM 64-QAM modulation with a block length $l = 108$ to align with the conditions described in~\cite{gultekin2021kurtosis,gultekin2022mitigating}. We also consider the FEC-independent sequence selection proposed in~\cite{civelli2023sequence} and ESS~\cite{gultekin2019enumerative} for distribution matching. We also use a 4D mapping strategy~\cite{skvortcov2021huffman} of our shaped amplitudes which has usually shown better performance than 1D and 2D mapping~\cite{askari2023probabilistic}. Our transmitted signal comprised $5$~wavelength-division multiplexing (WDM) channels, each operating at $50$~GBaud, resulting in a raw data rate of $600$~Gbit/s per channel. For all the tested ESS schemes, we employed a low-density parity-check (LDPC) code with a length of $64800$ bits, following the DVB-S2 standard, and a code rate $r_c = 5/6$. Consequently, the achieved net bit rate amounted to $8$~bits per 4D symbol, resulting in a net bit rate of $400$~Gbit/s per channel. The channel spacing is set to $55$~GHz, and we apply root-raised cosine (RRC) pulse shaping with a roll-off factor of $0.1$ to each channel. SSMF is simulated with an attenuation coefficient $\alpha_{\mathrm{dB}} = 0.2~\mathrm{dB/km}$, chromatic dispersion coefficient $D = 17~\mathrm{ps/nm/km}$, polarization mode dispersion (PMD) of $0.04~\mathrm{ps/\sqrt{km}}$, and a nonlinear parameter $\gamma = 1.3~\mathrm{(W\cdot km)^{-1}}$ at $\lambda = 1550~\mathrm{nm}$. We do not add any laser phase noise in the simulation to simplify our analysis. Following propagation, the central channel underwent several signal processing steps, including optical filtering, matched filtering, chromatic dispersion compensation, and genie-aided multiple-input-multiple-output (MIMO) channel equalization. Subsequently, for carrier phase recovery (CPR), we applied a fully-data-aided phase filter with a window averaging over $64$ symbols to compensate for phase rotation induced by cross-phase modulation (XPM), as elaborated in~\cite{fehenberger2015compensation}. Finally, we measured the electric SNR denoted $\mathrm{SNR_{elec}}$ from the equalized constellations and computed the generalized mutual information (GMI) using Eq.~(10) from~\cite{Bocherer14}.

For the higher baud rate long-haul link scenario, we evaluate the potential advantages of our E-SS and D-SS schemes employing digital-multi-band (DMB) signals known for their superior resistance to Kerr effects compared to single-carrier schemes~\cite{cho2022shaping,lorences2022improving}. The baud rate in this scenario is set at $110$~Gbaud, with $8$~ digital subcarriers, first in a single-wavelength configuration, then in a 5 WDM-channel configuration. The link is made of $30\times80$~km SSMF spans with one $5$~dB-noise-figure EDFA amplifier per span. The achieved net data rate per channel is $880$~Gbit/s. The digital signal processing (DSP) blocks employed in this scenario are consistent with those detailed in the preceding paragraph.

\section{From EDI to D-EDI}\label{EDI_TO_D-EDI}

EDI is a sign-independent metric proposed in~\cite{wu2021temporal}. We denote it as $\Psi\left[\mathbf{X}\right]$ in Eq.~\eqref{eq:PSI} defined as the ratio of the variance to the mean of the windowed energies in the sequence $\mathbf{G}^{w}$ defined in Eq.~\eqref{eq:Gw}. For a given window length $w$ and a $2\times L_s$ sequence $\mathbf{X}$ defined in Eq.~\eqref{eq:X} as $L_s$ 2D complex-valued column vectors or equivalently two $L_s$-long row vectors, each element $G_{i}^{w}$ of $\mathbf{G}^{w}$  is computed as the sum of the energies of $w+1$ 2D symbols centered on $\mathbf{x}_{i}$ as shown in Eq.~\eqref{eq:G}. $\mathrm{Var[\cdot]}$ and $\mathrm{E[\cdot]}$ are the variance and expectation of $\cdot$ respectively.
 
\begin{equation}
    \Psi\left[\mathbf{X}\right] \overset{\Delta}{=} \frac{\mathrm{Var}\left [\mathbf{G}^{w} \right ]}{\mathrm{E}\left [\mathbf{G}^{w} \right ]}
    \label{eq:PSI}
\end{equation}

\begin{equation}
    \mathbf{G}^w=\left[G_{\frac{w}{2}+1}^{w},G_{\frac{w}{2}+2}^{w}, \cdots,G^w_i,\cdots, G_{L_{s}-\frac{w}{2}}^{w}\right]
    \label{eq:Gw}
\end{equation}

\begin{equation}
\mathbf{X}=\left[\mathbf{x}_{1},\mathbf{x}_{2},\cdots,\mathbf{x}_{L_s}\right]=\begin{bmatrix}
        \mathbf{x}_{\mathrm{pol_1}}\\
        \mathbf{x}_{\mathrm{pol_2}}
    \end{bmatrix}   
\label{eq:X}
\end{equation}

\begin{equation}
    G_{i}^{w} = \sum _{k=i-\frac{w}{2}}^{i+\frac{w}{2}}\Vert \mathbf{x}_{k}\Vert^{2},~~i:\frac{w}{2}+1\rightarrow L_s-\frac{w}{2}
    \label{eq:G}
\end{equation}

The optimal EDI averaging window length $w$ (where $w \geq 2$ and is an even number), expressed as a number of symbols, is directly linked to the two-sided channel memory, denoted $2M$ and given by $2M = 2\left \lfloor\pi|\beta_2|BR_sL\right \rceil$ where $\beta_2$ stands for the group velocity dispersion, $B$ is the channel spacing, $R_s$ is the symbol rate, $L$ is the transmission distance, and $ \left \lfloor \cdot \right \rceil $ denotes the rounding to the nearest integer. As observed in previous works~\cite{wu2021temporal} (specifically in their Fig.~9), the optimal window length $w$, defined as the length where the correlation between EDI and SNR is maximized, is typically less than $2M$.

To further capture the energy variations within a finite-length sequence propagating over a link, we introduce the EDI of dispersed sequences, abbreviated as D-EDI and defined as the average of EDI values computed at multiple points along a linear dispersive and lossless fiber. This is shown in Fig.~\ref{fig:DE-SS} and in Eq.~\eqref{eq:DEDI} in which D-EDI is denoted as $\Psi_{D}$ and $\mathbf{X}^{(N)}$ designates the symbol sequence that has propagated over $N\times L_{D}$ kilometers of an ideal dispersive fiber obtained  using Eq.~\eqref{eq:Xm} and~\eqref{eq:Dispersion_applier}, $\mathrm{m_{D}}$ is the maximum number of dispersion operations and $\Psi\left[\cdot\right]$ is the EDI operator defined in Eq.~\eqref{eq:PSI}.

\begin{equation}
    \Psi_{D} \overset{\Delta}{=} \frac{1}{m_{D}+1}\sum_{N=0}^{m_{D}}\Psi\left[ \mathbf{X}^{(N)}\right]
    \label{eq:DEDI}
\end{equation}

\begin{equation}
    \mathbf{X}^{(N)} =\mathfrak{D}_{D, N\times L_{D}}\left [ \mathbf{X} \right ]
    \label{eq:Xm}
\end{equation}

\begin{equation}
    \mathfrak{D}_{D, z}\left [ \mathbf{X} \right ]\overset{\Delta}{=} \begin{bmatrix}
        \mathfrak{F}^{-1}\left( \mathfrak{F}\left(\mathbf{x}_{\mathrm{pol_1}}\right)\odot\exp \left( j\frac{D\lambda^2 \pi \boldsymbol{\mathbf{\Delta f}} ^2 z}{c} \right) \right)\\ 
        \mathfrak{F}^{-1}\left( \mathfrak{F}\left(\mathbf{x}_{\mathrm{pol_2}}\right)\odot\exp \left( j\frac{D\lambda^2 \pi \boldsymbol{\mathbf{\Delta f}} ^2 z}{c} \right) \right)
    \end{bmatrix}
    \label{eq:Dispersion_applier}
\end{equation}

The dispersion operator in Eq.~\eqref{eq:Dispersion_applier} consists in a chromatic dispersion filter applied in the frequency domain and capturing the dispersion of a fiber with dispersion coefficient $D$ and length $z$~km. $\mathfrak{F}$ and $\mathfrak{F}^{-1}$ are the fast Fourier transform (FFT) and inverse fast Fourier transform respectively, $\odot$ the element-wise multiplication, $j=\sqrt{-1}$ the imaginary unit, $\lambda$ the central wavelength of the signal, $c$ the speed of light in vacuum and $\mathbf{\Delta f}$ the vector of frequency bins over which the FFT is computed. Dispersion is digitally applied over the perfectly-synchronized unfiltered symbol sequence sampled at one sample per symbol. We clearly see that, unlike EDI, D-EDI requires the knowledge of sign bits to compute the form of a dispersed sequence. EDI is its special case when $m_D = 0$.

In a single-span link, the signal power is substantially high at the start of the span, leading to significant NLI. To capture waveform variations within the first few kilometers over which the power loss is relatively minimal, we can compute the EDI multiple times shortly after the transmission begins within this span to derive the D-EDI.
In what follows, to keep the computational complexity low, for a single-span link, we add up two EDIs, the one computed over the sequences at the transmitter output and the one computed after propagation over the effective length of the fiber span $L_{D\mathrm{(single~span)}} = L_\mathrm{{eff}} = \frac{1-\mathrm{exp}(-\alpha L)}{\alpha}$~\cite{agrawal2000nonlinear} where $\alpha$ is the fiber attenuation in $\textrm{km}^{-1}$ and $L$ is the length of the fiber span in km. 

For a multi-span link consisting of $m$ spans of $L$ kilometers each, we may compute EDI several times within each span to capture the evolution of the waveform distortion. In this paper, to limit the complexity of the metric computation, we calculate the EDI of the sequence at the beginning of each span and combine them to obtain the D-EDI, hence $m_D = m-1$ and $ L_D = L$. We may reduce the complexity by computing EDIs less often, i.e. reducing $m_D$ and increasing $L_D$ as shown in Sec.~\ref{Sec: Complexity}.

\section{Implementation of EDI and D-EDI: E-SS and D-SS design}\label{E-SS and D-SS design}

\begin{figure*}[t]
    \centering
    \includegraphics[width=1\linewidth]
    {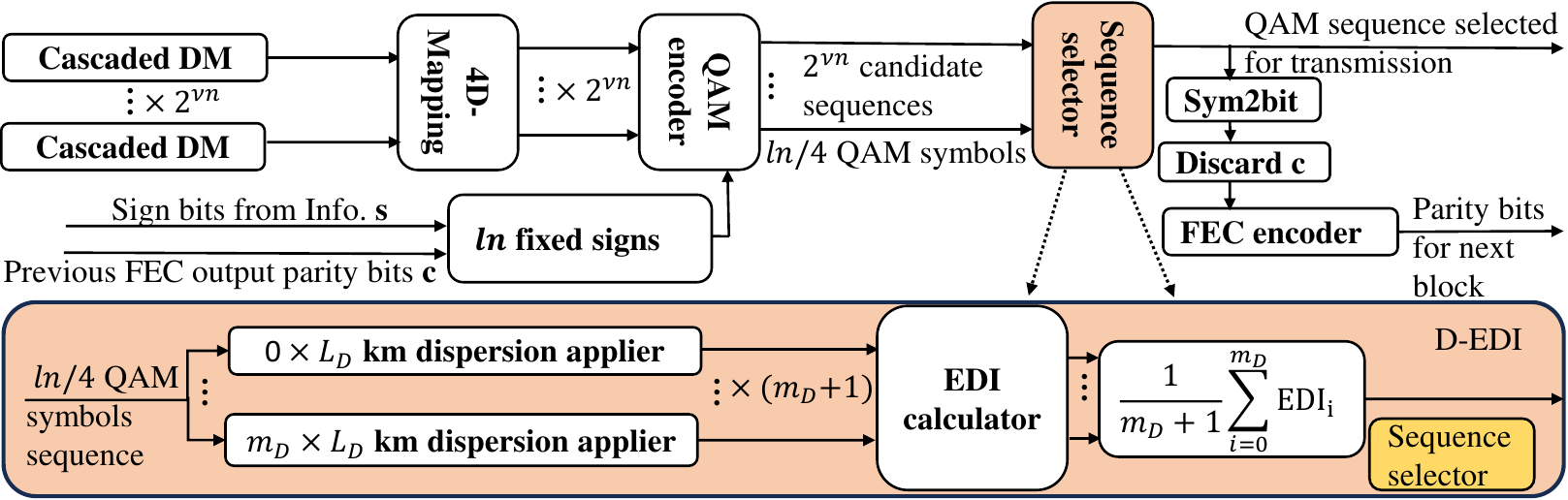}
    \caption{Block diagram of FEC independent D-SS in the PAS transmitter: EDI of dispersed sequences (D-EDI) is determined by averaging the calculated EDI at multiple locations along an ideal dispersive fiber. The dispersion is applied at $1$ sample per symbol. EDI is a special case of D-EDI where $m_D = 0$. The sign bits are fixed through a multi-block FEC-independent sequence selection process as shown in~\cite{civelli2023sequence}.}
    \label{fig:DE-SS}
\end{figure*}

\begin{figure}
    \centering
    \includegraphics[width=1\linewidth]
    {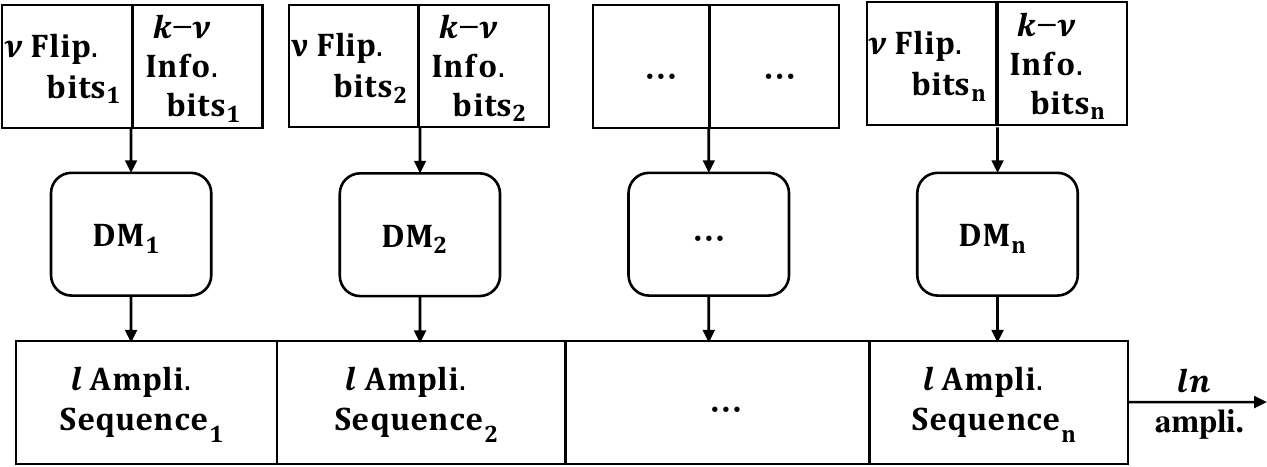}
    \caption{Structure of cascaded distribution matcher (DM). The output amplitude sequences of $n$ DMs are concatenated to form a longer sequence.}
    \label{fig:serial_DM}
\end{figure}

The fundamental concept behind E-SS and D-SS draws inspiration from the list-CCDM~\cite{wu2022list}. Our approach extends the concept for the use of various distribution matchers (DMs) and we concatenate the output of several DMs to generate a longer candidate sequence. In this study, we only use ESS for distribution matching.

In Fig.~\ref{fig:DE-SS}, we plot the encoding process where input bits are fed into a 1D ESS distribution matcher (DM) with block length $l$ that generates $l$ amplitudes. The input bit stream is composed of two parts: $k - \nu$ information bits and $\nu$ prefix flipping bits. Altering these $\nu$ prefix flipping bits can cause pronounced changes in the indices of the sequences during the ESS encoding process. Such changes are reflected as considerable alterations in the amplitudes across the block. Conveniently, these flipping bits are discarded after the decoding process at the receiver side. The outputs from $n$ 1D-DM can be sequentially concatenated, from head-to-tail, to create extended sequences, which can then undergo a selection process as shown in Fig.~\ref{fig:serial_DM}. This resulting sequence will have a length of $ ln $. Cascading DMs have two advantages: they permit obtaining more candidate sequences while maintaining the same rate loss, and at the same time, we maintain the use of short-block-length DMs that showed a clear advantage against long-block-length DM with respect to nonlinear effects. For an E-SS (also true for D-SS) with $n$ cascaded DMs, where each DM includes $\nu$ flipping bits, we can generate $2^{\nu n}$ different candidate sequences. Subsequently, the $ln$ amplitudes are mapped to $L_{s} = ln/4$ 4D-QAM symbols after 4D-mapping~\cite{skvortcov2021huffman}. 
 
E-SS employs EDI as a metric for sequence selection, hence the assigned signs to the generated amplitudes are not necessary for the evaluation of EDI. The sequence selector then identifies the $2^{\nu n}$ different candidate sequences and transmits the sequence with the lowest 4D-energy EDI.

D-SS employs D-EDI as a metric for sequence selection, which requires the assigned signs for the metric evaluation and selection. The sign bits can be fixed through a multi-block FEC-independent sequence selection process as proposed in~\cite{civelli2023sequence}. In this scenario, the sign bits are chosen in two parts: a part from the information stream of the current D-SS block, denoted~\textbf{s}, and a part from the FEC output parity bits of the previous D-SS block, denoted~\textbf{c} as shown in Fig.~\ref{fig:DE-SS}. Once the QAM sequence is selected, it is demapped into a bit sequence and sent into an FEC encoder, after discarding the parity bits from the previous D-SS block \textbf{c}, to generate parity bits for the next D-SS block.


The sequence selection process introduces an additional rate loss~\cite{civelli2021sequence,secondini2022new},  which is $\frac{\nu}{l}$ accounting for the effect of the flipping bits as in~\cite{askari2023probabilistic}. The total rate loss $R_\mathrm{{loss}}$ of E-SS and D-SS in each dimension is:

\begin{equation}
    R_{\mathrm{loss}} = H(P_{a})-\frac{k-\nu}{l}.
    \label{eq:R_loss}
\end{equation}
where $H(P_{a})$ is the entropy linked to the distribution $P_{a}$ and $P_{a}$ is the marginal distribution of amplitudes of 1D PAM symbols. As $\nu$ increases, the number of candidate sequences also increases, resulting in higher nonlinear shaping gain. However, this comes at the cost of an increased rate loss. As a result, an efficient sequence selection mechanism should strike a balance between linear and nonlinear shaping gains.

\begin{figure}[t]
    \centering
    \includegraphics[width=1\linewidth]
    {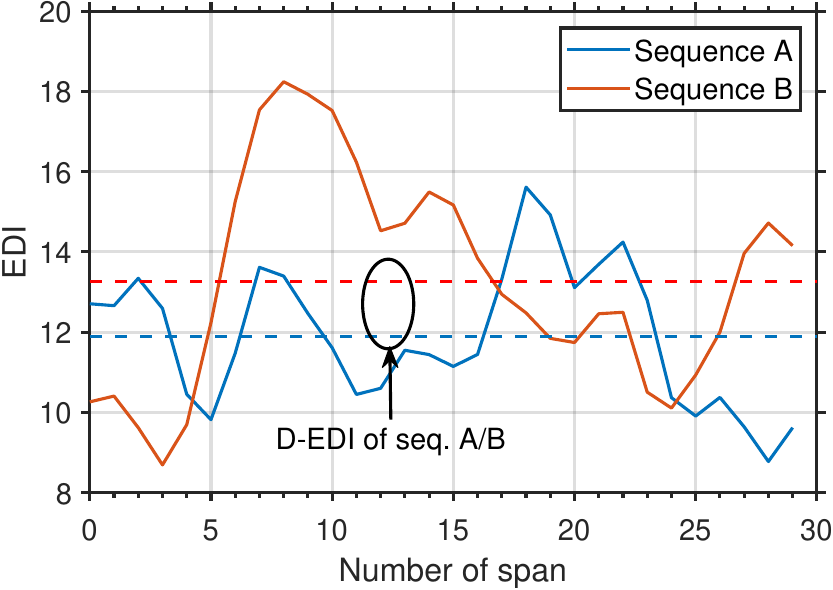}
    \caption{Simulation in single channel, $8$~DMB subcarriers, $110$~GBaud transmission over $30\times80$~km SSMF link, with a net rate of $880$~Gbit/s. Sequences A and B are two random sequences generated by single ESS with block length $l = 108$ over the $4^{\mathrm{th}}$ subcarrier (i.e. central subcarrier), each consisting of $108$ 4D 64-QAM symbols. Solid lines represent the evolution of EDI at the beginning of each span, dashed lines represent D-EDI computed as an averaged EDI over all spans. D-EDI gives a more accurate estimation of NLI generated by each sequence after transmission.}
    \label{fig:EDI_DEDI_2sequence}
\end{figure}

In Fig.~\ref{fig:EDI_DEDI_2sequence}, we illustrate the evolution of the EDI metric at the beginning of each span when transmitting two different random sequences over a $30\times80$~km link with a single channel, $8$~DMB subcarriers, $110$~GBaud transmission, and a net rate of $880$~Gbit/s. Each of these sequences comprises $108$ random 4D 64-QAM symbols over the $4^{\mathrm{th}}$ subcarrier. EDI is calculated at the beginning of each span and both EDI and D-EDI employ a window length of 2 symbols ($w
= 2$), that is $1$ symbol before and $1$ symbol after the current symbol. Before transmission, sequence B exhibits a lower EDI compared to sequence A. This implies that sequence B will generate less NLI over the first span than sequence A. However, as the sequences propagate through the dispersive fiber, their characteristics change continuously due to the influence of dispersion. After only $4$ spans, the EDI of sequence B surpasses that of sequence A. Ultimately, when the D-EDI is computed by averaging EDI over all spans, sequence A yields a smaller D-EDI even though it had a larger initial EDI. This suggests that sequence A maintains a higher average nonlinear shaping gain throughout the entire transmission. This outcome emphasizes the importance of considering sign-dependent metrics, such as D-EDI, when assessing the nonlinear shaping gain performance of sequences. Sign-independent metrics may not provide a complete picture of NLI manifestation, as the influence of dispersion can cause initially favorable sequences to show deteriorated performance after a certain propagation distance, while initially unfavorable sequences may show an improved performance. Hence, D-EDI allows for a more accurate estimation of NLI generated by each sequence throughout a transmission.

\section{Performance assessment: Single-span transmission}
\label{Single span transmission}


\begin{figure*}[htb]
    \centering
    \includegraphics[width=1\linewidth]
    {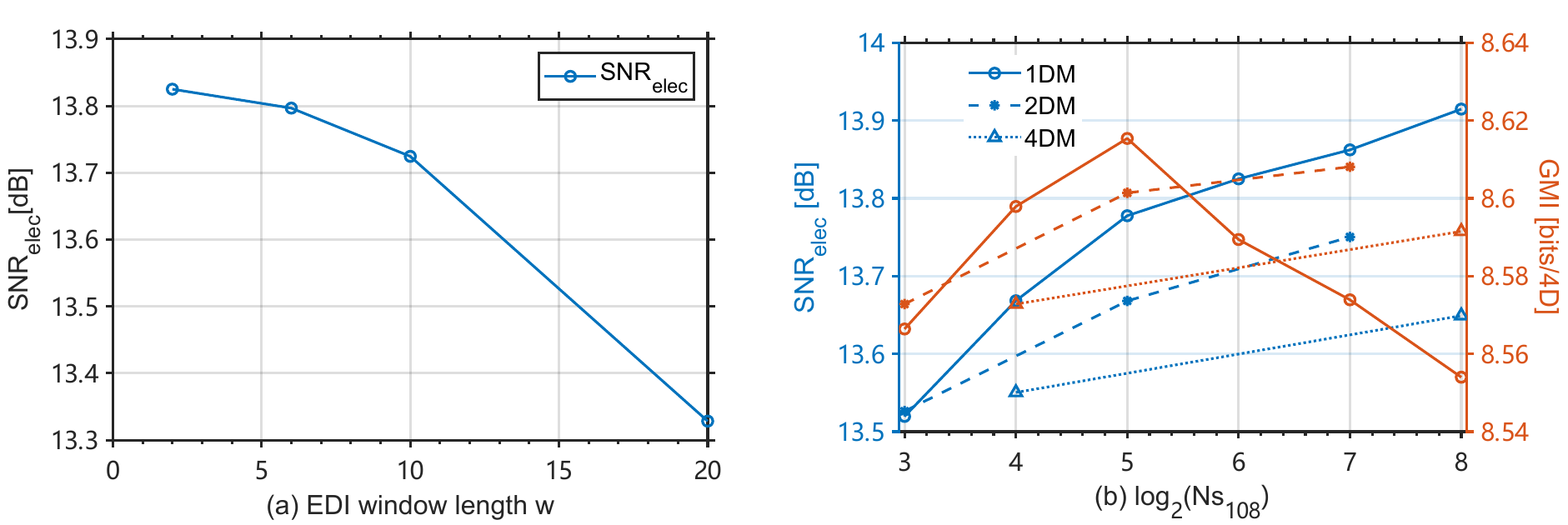}
    \caption{Simulation of $50$~GBaud E-SS transmission over $205$~km single span SSMF link, $5$~WDM channels, $55$~GHz channel spacing, net rate per channel $400$~Gbit/s. (a): Electrical SNR versus the EDI window length $w$ used in the sequence selector to evaluate the EDI of sequences. $\nu=3$ flipping bits and a single 1D-DM ($n=1$) is used. The power per channel is $9$~dBm. (b): Electrical SNR and GMI at $9$~dBm versus $\log_{2}(N_{S_{108}})$ when cascading 1, 2 or 4 DMs, where $\mathrm{N_{S_{108}}}$ is the number of required sequence selections to generate $108$~4D-symbols. $\mathrm{E\text{-}SS_{1}^{3}}$ exhibits the highest performance in terms of GMI (ESS without SS exhibits $13.27$~dB electrical SNR value and $8.46$~bits/4D GMI value).} 
    \label{fig:Optimization_205km}
\end{figure*}

\begin{figure*}[htb]
    \centering
    \includegraphics[width=1\linewidth]
    {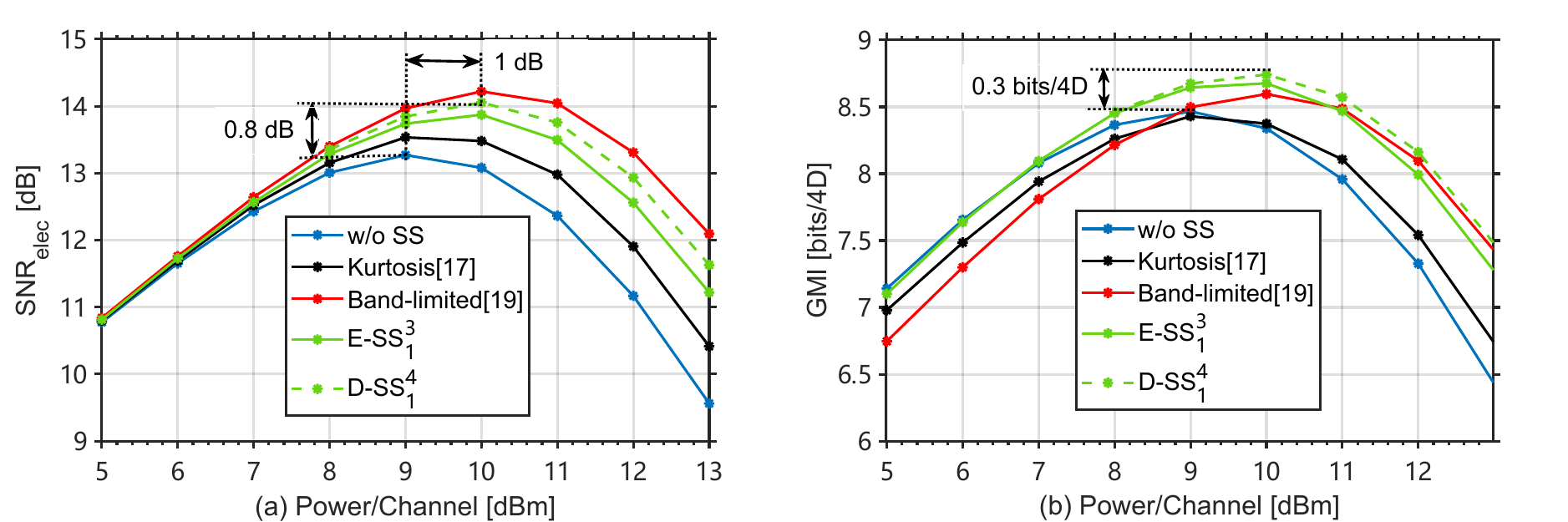}
    \caption{Simulation of ESS without SS, K-ESS, B-ESS, $\mathrm{E\text{-}SS_{1}^{3}}$, and $\mathrm{D\text{-}SS_{1}^{4}}$ (best-performing scheme) with D-EDI computed after applying a single $21$~km dispersion filter. Same system configuration as for Fig.~\ref{fig:Optimization_205km}. Left: Electrical SNR versus power per channel. Right: GMI versus power per channel. Both $\mathrm{E\text{-}SS_{1}^{3}}$ and $\mathrm{D\text{-}SS_{1}^{4}}$ demonstrate superior performance when compared to state-of-the art schemes.}
    \label{fig:Performance_205km}
\end{figure*}

The single-span transmission scenario was introduced in Sec.~\ref{System_Description}. First, we focus on the E-SS scheme. In Fig.~\ref{fig:Optimization_205km}~(a), we plot the $\mathrm{SNR_{elec}}$ for various EDI window lengths $w$ used by the sequence selector. For this test, we used a single 1D-DM ($n=1$) and $\nu=3$ flipping bits and denote the scheme $\mathrm{E\text{-}SS_{1}^{3}}$ (or in general $\mathrm{E\text{-}SS_{n}^{\nu}}$). This setup allows us to select the most suitable sequence from eight possible candidates ($2^3=8$). We calculate $\mathrm{SNR_{elec}}$ using equalized symbols after transmitting over a single span with a channel power of $9$~dBm, which is near the optimal power level. Our findings show that as $w$ increases, the $\mathrm{SNR_{elec}}$ gradually declines. We find that a window length $w = 2$ yields the best performance, and is much lower than the channel memory ($2M=76$ when considering the total length of the span) as the NLI is generated mostly at the beginning of the single-span transmission. Therefore, we proceed with $w = 2$ in subsequent single-span simulations. 

In Fig.~\ref{fig:Optimization_205km}~(b), we analyze the performance of E-SS by varying the number of sequentially cascaded DMs $n$, where each DM has $\nu$ flipping bits. A single DM with a block length $l=108$ produces $27$ 4D-symbols after 4D-mapping. When cascading $ n = 1,2$ or $4$ DMs, the candidate sequence lengths will be $ 27, 54$ and $ 108 $ respectively. As a complexity metric, we designate the number of tested candidate sequences per $4$ DMs, i.e. per $108$ 4D-symbols as $\mathrm{N_{S_{108}}}$. We choose $4$ DMs as $4$ is the least common multiple between the three studied values of $n=\left\lbrace1,2,4\right\rbrace$. To achieve a total sequence of $108$ 4D symbols, the E-SS transmitter needs to be called $\frac{4}{n}$ times, resulting in $4, 2 $ and $ 1 $ calls for $ n = 1, 2, $ and $ 4 $ cascaded DMs respectively. Hence, $\mathrm{N_{S_{108}}}$ is given by:

\begin{equation}
    N_{S_{108}} = \frac{4}{n}2^{\nu n}
\end{equation}

We see from Fig.~\ref{fig:Optimization_205km}~(b) that, as $\mathrm{N_{S_{108}}}$ increases, $\mathrm{SNR_{elec}}$ also increases for $n=\left\{ 1,2,4\right\}$. However, as the number of flipping bits increases, the rate loss likewise increases. The maximum GMI is hence attained when achieving a trade-off between rate loss and nonlinear gain. Specifically, the largest GMI is achieved with a single 1D-DM at $\log_{2}(N_{S_{108}}) = 5$ ($32$ candidate sequences), corresponding to the $\mathrm{E\text{-}SS_{1}^{3}}$ configuration. 

Next, we also evaluated the performance of the D-SS scheme over the same link by computing D-EDI as the sum of the EDI of the sequence at the transmitter output and its EDI after propagating over a linear dispersive fiber without attenuation of length $L_\mathrm{{eff}}$ over which nonlinear effects remain significant. $n=1$ and $\nu=4$~($\mathrm{D\text{-}SS_{1}^{4}}$) yields the best GMI gains through a similar optimization as the one shown in Fig.~\ref{fig:Optimization_205km}(b) for E-SS. In Fig.~\ref{fig:Performance_205km}~(a), we show $\mathrm{SNR_{elec}}$ versus the launch power per channel for five different ESS schemes. The D-SS, E-SS, band-limited ESS (B-ESS), and kurtosis-limited ESS (K-ESS) exhibit SNR gains of approximately $0.8$~dB, $0.6$~dB, $1$~dB, and $0.3$~dB respectively, compared to the ESS without sequence selection (SS) at a power level of $10$~dBm. Additionally, they demonstrate up to $1$~dB improvement in the optimal launch power. In Fig.~\ref{fig:Performance_205km}~(b), we present the results in terms of GMI per 4D symbol. Notably, the D-SS scheme yields the highest GMI. Specifically, at a power level of $10$~dBm, it achieves a GMI increase of $0.15$~bits/4D compared to B-ESS and a $0.3$~bits/4D improvement compared to ESS without SS. The E-SS scheme also delivers a GMI gain of $0.1$~bits/4D (respectively $0.2$~bits/4D) in comparison to B-ESS (respectively ESS without SS). It is important to mention that both EDI and D-EDI based shaping approaches exhibit nearly the same linear performance as unconstrained ESS due to their small rate loss. E-SS and D-SS exhibit similar performance largely because the initial shape of the waveform at the start of the link plays a crucial role in the generation of NLI for single span links. However, a slight performance improvement is achievable by selecting sequences that also exhibit low EDI after the effective length $L_\mathrm{{eff}}$. In terms of rate loss, D-SS demonstrates superiority over B-ESS and K-ESS, with a rate loss of $0.26$~bits/4D compared to $0.5$~bits/4D and $0.27$~bits/4D respectively. The rate losses are not discernible in Fig.~\ref{fig:Performance_205km}~(b) between $5$ dBm and the optimal power level because nonlinear gains are already occurring within these power ranges. D-SS also outperforms E-SS in the nonlinear regime. Overall, it outperforms B-ESS in both linear and nonlinear regimes. Finally, we remind that D-SS requires a slightly higher complexity compared to E-SS since it necessitates knowledge of sign bits and computation of dispersed sequences at the transmitter side.

\section{Performance assessment: Multi-span transmission}
\label{Multi span transmission}


\subsection{Optimization and Transmission Performance}

\begin{figure}[htb]
    \centering
    \includegraphics[width=1\linewidth]
    {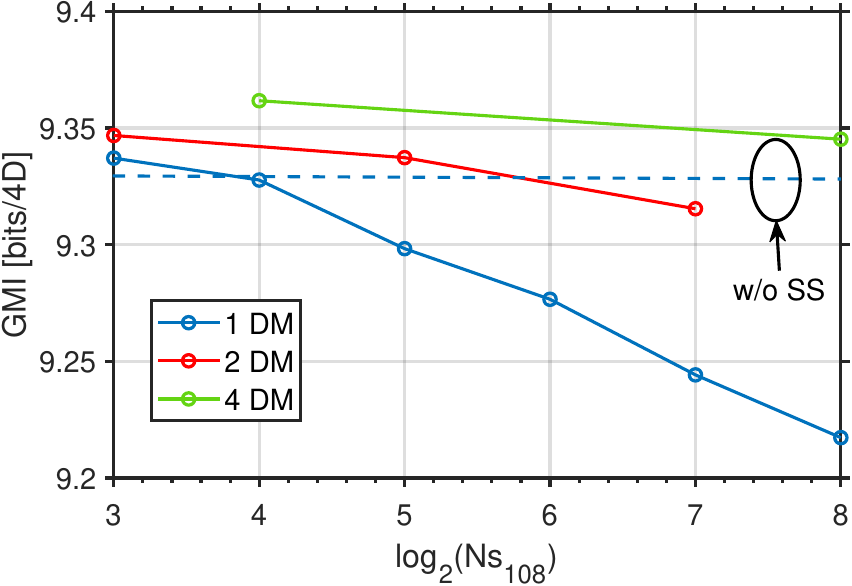}
    \caption{Simulation of D-SS with D-EDI computed after applying dispersion successively over $30\times80$~km, single channel, $8$~DMB subcarriers, $13.75$~GBaud per subcarrier transmission, net rate per channel $880$~Gbit/s. GMI versus $\log_{2}(N_{S_{108}})$ at the optimal power ($4.5$~dBm), where $\mathrm{N_{S_{108}}}$ is the number of tested sequences per $108~4$D-symbols ($w=2$ for all multi-span test, corresponds to the memory of a $13.75$~Gbaud transmission over a single $80$~km span). $\mathrm{D\text{-}SS_{4}^{1}}$ exhibits the best performance.}
    \label{fig:SNR_elec_GMI_vs_complexity}
\end{figure}

The multi-span transmission scenario was introduced in Sec.~\ref{System_Description}. For the DMB scheme, we select sequences for a single digital band and then distribute them over the eight bands. Performing sequence selection jointly for all digital bands would introduce significant complexity that we avoid in this work. In Fig.~\ref{fig:SNR_elec_GMI_vs_complexity}, we evaluate the performance of our D-SS scheme for different cascaded DMs and different numbers of flipping bits. In long-distance links, accumulated chromatic dispersion becomes apparent as its induced inter-symbol interference involves hundreds of symbols. For these links, it becomes more important to cascade multiple 1D-DMs to generate extended candidate sequences over which we will compute D-EDI. By doing so, we account for the impact of the accumulated channel memory on the waveform, thus making a better selection process. In addition, for the same complexity level as a single DM, this approach helps in reducing rate loss. As we did in the previous section, the complexity of the scheme is evaluated as $\mathrm{N_{S_{108}}}$, the number of tested candidate sequences for the generation of $108$~4D-symbols. We calculate D-EDI by summing up EDI values of the sequences at the input of each of the $30$ spans constituting the $30\times80$~km link ($w=2$ corresponding to the memory of a $13.75$~Gbaud transmission over a single $80$~km span). We observe that the largest GMI occurs in the configuration where four 1D-DMs are arranged in series, and each 4D-DM has one flipping bit, denoted $\mathrm{D\text{-}SS_{4}^{1}}$, and corresponding to $\log_ {2}(N_{S_{108}}) = 4$ (i.e. $16$ candidate sequences). The GMI curve for a single DM exhibits a significantly steeper slope compared to the one for four DMs, as the rate loss increases at a slower pace in the case of cascading four DMs.

\begin{figure}[tb]
    \centering
    \includegraphics[width=1\linewidth]
    {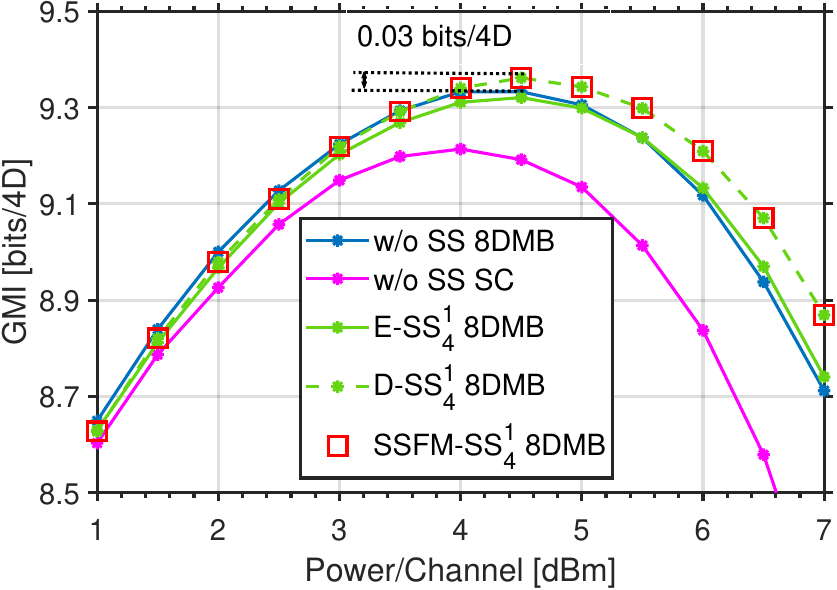}
    \caption{Simulation of ESS without SS, $\mathrm{E\text{-}SS_{4}^{1}}$, $\mathrm{D\text{-}SS_{4}^{1}}$ and $\mathrm{SSFM\text{-}SS_{4}^{1}}$ with $30\times80$~km dispersion applier over $1$~channel $8$~DMB subcarriers, and ESS without SS with $1$~channel $1$~carrier, in $110$~GBaud over $30\times80$~km SSMF link, net rate per channel $880$~Gbit/s. Electrical SNR versus Power per channel in the non-linear channel. $\mathrm{D\text{-}SS_{4}^{1}}$ exhibits the same performance as that of an ideal SSFM-based sequence selection.}
    \label{fig:Performance_30spans}
\end{figure}

Next, in Fig.~\ref{fig:Performance_30spans}, we compare the performance of four single-wavelength transmission schemes: ESS without SS over a single carrier (SC) serving as a baseline, ESS without SS over DMB using $8$ subcarriers (8DMB), $\mathrm{D\text{-}SS_{4}^{1}}$ (optimal case) and $\mathrm{E\text{-}SS_{4}^{1}}$ schemes, both in 8DMB mode. The $\mathrm{D\text{-}SS_{4}^{1}}$ shows a $0.03$~bits/4D-symbol gain in GMI at a power of $4.5$~dBm compared to ESS without SS and 8DMB. It also shows a $0.16$~bits/4D in GMI at the optimal power compared to ESS without SS over a single carrier. On the other hand, the $\mathrm{E\text{-}SS_{4}^{1}}$ has no or very limited gain in GMI. Hence, the sign-dependent-based D-SS scheme maintains the GMI advantage in both short-distance and long-haul distance transmission scenarios. 

Additionally, we conducted simulations involving sequence selection based on the computation of the actual NLI produced by each sequence using the Split-step Fourier method (SSFM sequence selection or SSFM-SS) to approximate an upper bound for the performance of a sequence selection scheme. We kept the transmitter structure as 4 concatenated DMs, each equipped with $1$ flipping bit. This approach is similar to the one outlined in~\cite{civelli2023sequence}. Each candidate sequence was transmitted over a $30\times80$~km standard single mode fiber, single-wavelength, single-carrier system operating at $13.75$~Gbaud, without the addition of ASE noise. After transmission, we compare the NLI power of candidate sequences and select the sequence that generates the lowest NLI. This process allows us to obtain the optimal performance of sequence selection with $4$ serial DMs, and $1$ flipping bit when only considering self-phase modulation effects. From Fig.~\ref{fig:Performance_30spans}, we observe that SSFM-SS and D-SS exhibit almost identical performance. This observation indicates that within the analyzed system, selecting sequences based on D-EDI can be a viable alternative to the more complex SSFM approach. In other words, the variations of the waveform shaped by dispersion effects primarily dictate nonlinear distortions in high-power regions. Nonlinearity, in this context, acts as a small perturbation. Consequently, we can feasibly substitute SSFM with the dispersion-aware metric D-EDI to perform sequence selection with reduced complexity.

\begin{figure*}[htb]
    \centering
    \includegraphics[width=1\linewidth]
    {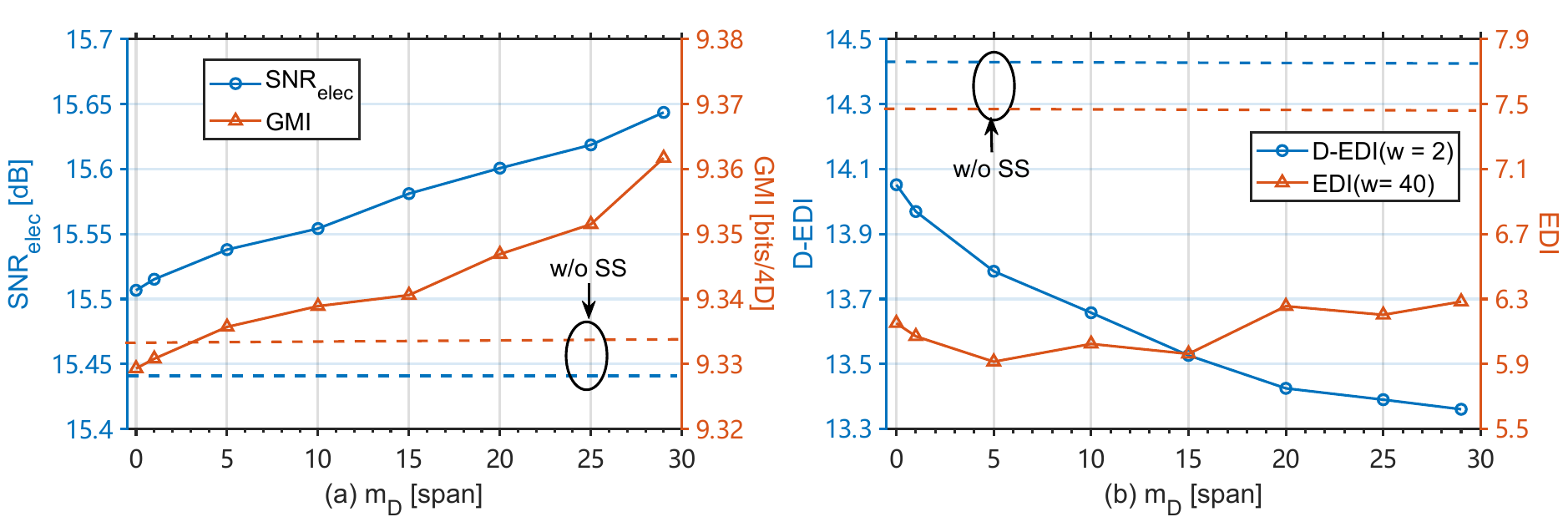}
    \caption{Performance analysis of $\mathrm{D\text{-}SS_{1}^{4}}$ (a): GMI versus $\mathrm{m_{D}}$ at optimal power ($4.5$~dBm) where $\mathrm{m_{D}}$ is the maximum number of spans used by the dispersion applier during sequence selection. Gains are maintained even when we reduce the complexity by computing D-EDI over fewer spans. (b): D-EDI and EDI versus $\mathrm{m_{D}}$.  We add the values for ESS without SS by horizontal dashed lines. Same system configuration as in Fig.\ref{fig:SNR_elec_GMI_vs_complexity}. D-EDI demonstrates a negative correlation with $\mathrm{SNR_{elec}}$.}
    \label{fig:Performance_of_m_span_max}
\end{figure*}

\subsection{Complexity Reduction and Interpretation of the Results} \label{Sec: Complexity}
\begin{figure}[t]
    \centering
    \includegraphics[width=1\linewidth]
    {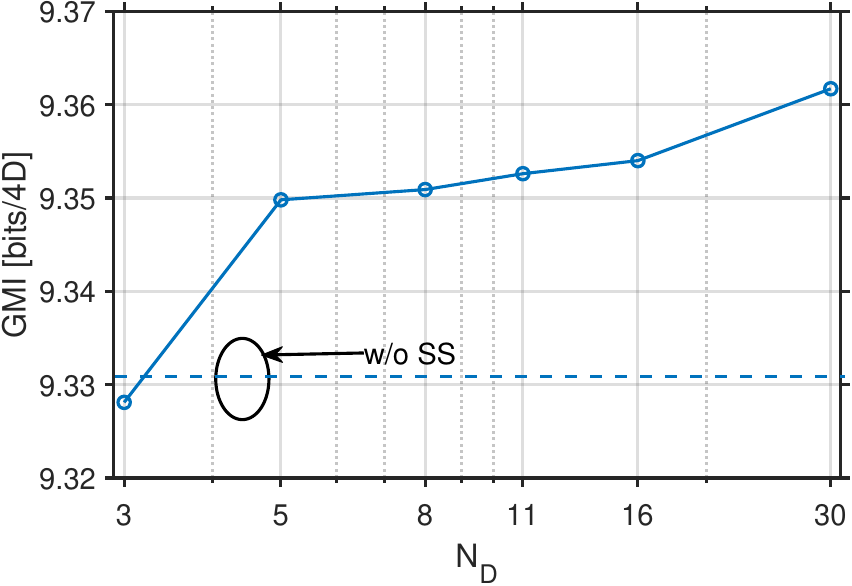}
    \caption{GMI versus $N_{D}$ for $\mathrm{D\text{-}SS_{1}^{4}}$, where $N_{D}$ is the number of EDI calculations required to compute D-EDI, each $N_{D}$ value represents the following span indices: $N_{D}= 3\rightarrow[0,1,29]; 5\rightarrow[0,1,9:10:29]; 8\rightarrow[0,1,4:5:29]; 11\rightarrow[0,1,2:3:29]; 16\rightarrow[0,1:2:29]; 30\rightarrow[0:1:29]$. We add the values for ESS without SS by horizontal dashed lines.  Same system configuration as in Fig.~\ref{fig:SNR_elec_GMI_vs_complexity}. Gains are maintained even when we reduce the complexity by computing D-EDI over several spans.}
    \label{fig:Performance_of_L_segment}
\end{figure}

In the above study, we aimed to improve sequence selection by calculating the EDI at the beginning of each span along the entire transmission link. This resulted in obtaining $30$ different EDI values averaged to derive the final D-EDI. Now, we delve into the impact of reducing the number of EDI computations on the achieved performance. To do so, we explore two methods. First, we consider calculating D-EDI for only the first $m_{D}=\left\{1,\cdots,29\right\}$ spans instead of the entire transmission line. It is worth reminding that when $m_{D} = 0$, D-EDI simplifies to the original EDI. Second, we consider limiting the number of EDI calculations to $N_{D}<30$ by computing EDI only at the beginning of specific spans described in the caption of Fig.~\ref{fig:Performance_of_L_segment}, instead of computing it at the beginning of each span ($N_{D} = 30$).

In Fig.~\ref{fig:Performance_of_m_span_max}, we explore the first method. In Fig.~\ref{fig:Performance_of_m_span_max}~(a), we present the performance results in terms of $\mathrm{SNR_{elec}}$ and GMI. We include the values for ESS without SS as horizontal dashed lines for comparison. SNR exhibits a steady growth as $\mathrm{m_{D}}$ increases. The peak value is achieved when $m_{D} = 29$, implying that D-EDI calculations across the entire link yield the maximum benefit. After only three spans, the nonlinear gain brought by D-EDI-based sequence selection overcomes the penalty in increased rate loss, resulting in a higher GMI than the one obtained with ESS without SS. In dispersion unmanaged transmission systems, as analyzed by the GN~\cite{poggiolini2013gn} and EGN models~\cite{carena2014egn}, nonlinear penalties are equally distributed and independent across homogeneous spans. Consequently, as $\mathrm{m_{D}}$ increases, the calculation for D-EDI encompasses an increasingly larger number of spans. This inclusion progressively mitigates the NLI across all spans. As a result, the system performance improves linearly with the increase of $\mathrm{m_{D}}$. Hence, to get the most out of sequence selection, it is crucial to calculate D-EDI for the entire link. Furthermore, it is important to note that the calculation of D-EDI only accounts for the influence of dispersion. However, in actual optical transmission scenarios, as the transmission distance increases, NLI also escalates, subsequently impacting energy fluctuations. Consequently, the accuracy of D-EDI calculations will be generally higher when computed over a specific number of initial spans compared to a computation over an equivalent number of final spans. This higher accuracy in the early stages of transmission translates into better performance improvements following sequence selection. 

Thereafter, in Fig.~\ref{fig:Performance_of_m_span_max}~(b), we illustrate the EDI and D-EDI values evaluated for the selected candidate sequences for different $\mathrm{m_{D}}$ values. EDI is evaluated over $40$~symbols ($w = 40$), which is an empirical value lower than the total channel memory $2M=68$ (corresponding to the channel memory of a $13.75$~Gbaud transmission over the entire link ($30\times80$~km)) as explained in Sec.~\ref{E-SS and D-SS design}, whereas D-EDI is assessed by averaging EDIs computed with a memory of $2$~symbols ($w = 2$), corresponding to the memory of a $13.75$~Gbaud transmission over a single $80$~km span. This D-EDI used for performance evaluation of the selected sequences is calculated along the entire link. Notably, as $\mathrm{m_{D}}$ increases, D-EDI continues to decrease and hence exhibits a negative correlation with $\mathrm{SNR_{elec}}$, while EDI struggles to do so, showing that D-EDI is a better estimator of the performance of the selected sequences for multi-span transmissions.

Next, we explore the second reduced-complexity D-EDI computation method. Instead of computing EDI at the start of each span, we choose to compute it for fewer spans specified in the caption of Fig.~\ref{fig:Performance_of_L_segment}. The figure shows the performance variations in terms of GMI versus the number of required EDI calculations $N_{D}$. We observe that, with decreasing $N_{D}$, GMI experiences a gradual decline. Significantly, even with $N_{D}=5$ (corresponding to EDI calculations every $800$ km), the system still outperforms ESS without SS. In this case, the computational complexity of D-EDI is significantly reduced compared to $N_{D} = 30$. This finding suggests that, for digital multi-band (DMB) transmissions over long-haul links, we can still achieve gains while using a metric with a reduced computational complexity for sequence selection.

To conclude this subsection, we highlight two key observations from the analyses in Fig.~\ref{fig:Performance_of_m_span_max} and Fig.~\ref{fig:Performance_of_L_segment} for DMB transmissions over long-haul links: first, a sequence that performs well for D-EDI over fewer spans may not necessarily maintain its effectiveness over longer links. This observation underscores the complexity of sequence selection in optical communications. It is not sufficient to assess the suitability of a sequence based solely on its performance at a single point in the link, such as the beginning of a link or elsewhere, or at a part of the total distance, which is also proved by Fig.~\ref{fig:EDI_DEDI_2sequence}, in which the EDI for two random sequences exhibits continuous variations over the transmission distance. Second, identifying an optimal sequence does not necessarily require an exhaustive analysis at the start of every span. Although dispersion-induced interference accumulates over longer distances, causing significant power variations and nonlinear phase fluctuations, our goal is to identify sequences that has a lower D-EDI over the entire distance. 
Since interference typically builds up over an extended period, involving the overlap of several symbols, it is practical to evaluate sequences at intervals spanning several spans rather than continuously or for every span. 

\subsection{The Impact of Block Length and Sequence Length Variations}
Up to now, we have used a fixed block length of $108$ for the comparison of our proposed schemes with state-of-the-art schemes in similar conditions. It is important to note that in the case of ESS, the block length has a significant impact on the performance (GMI). Decreasing the block length leads to increased nonlinear gain but also results in greater rate loss. Consequently, the optimal block length for ESS strikes a balance between maximizing nonlinear gain and minimizing rate loss. In Fig.~\ref{fig:performance_vs_BL}, we plot the performance of ESS schemes for four different block lengths $l~=\left\{60,108,200,300\right\}$. The considered schemes are ESS without SS alongside D-SS with $16$ candidate sequences and D-SS with $256$ candidate sequences. With the increase of cascaded DMs and flipping bits, the number of candidate sequences can rapidly increase beyond $16$ or $256$. For example, when $4$ serial DMs, each with $3$ flipping bits, are concatenated, there are $2^{4\times3} = 4096$ candidate sequences. To limit and maintain a relatively fixed complexity, we randomly select $16$ or $256$ sequences for selection to examine both low and high complexity schemes. In Fig.~\ref{fig:performance_vs_BL}, for each block length and each number of tested candidate sequences, we only show the best D-SS performance found after testing various configurations of cascaded DMs and flipping bits. The main observations are:

\begin{figure}[t]
    \centering
    \includegraphics[width=1\linewidth]
    {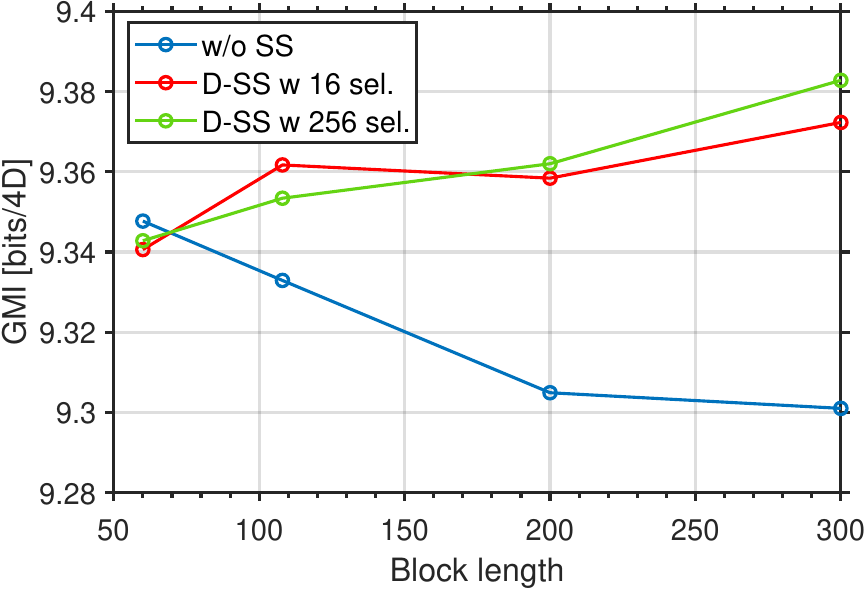}
    \caption{GMI versus DM block length. We show the optimal performance for each block length by changing the number of flipping bits and the number of cascaded DMs. Same system configuration as in Fig.\ref{fig:SNR_elec_GMI_vs_complexity}. D-SS demonstrates superior performance over different block lengths.}
    \label{fig:performance_vs_BL}
\end{figure}

\begin{itemize}

\item  For the ESS without SS, as the block length decreases, the rate loss gradually increases. Concurrently, the shortest block length of $60$ provides a nonlinear gain that exceeds the relatively high rate loss, resulting in the maximum achievable GMI.

\item  For D-SS, at a block length of $60$, the rate loss incurred by ESS itself, along with the additional rate loss from flipping bits, becomes too substantial for the achievable rate. Even though D-SS exhibits a larger nonlinear gain, the final GMI falls short of what ESS can offer.

\item  For longer block lengths, D-SS consistently outperforms ESS due to its ability to maintain a significant nonlinear gain while mitigating excessive rate loss. Among these block lengths, D-SS with a block length of $300$ achieves the highest GMI.

\item  In particular, when the block length is equal to $300$, D-SS with $256$ candidate sequences shows a $0.08$~bits/4D GMI advantage over ESS with the same block length and a $0.04$-bits/4D advantage over ESS with a block length of $60$. D-SS with $16$ selections shows similar trends with slightly lower gains.
\end{itemize}

\begin{figure}[t]
    \centering
    \includegraphics[width=1\linewidth]
    {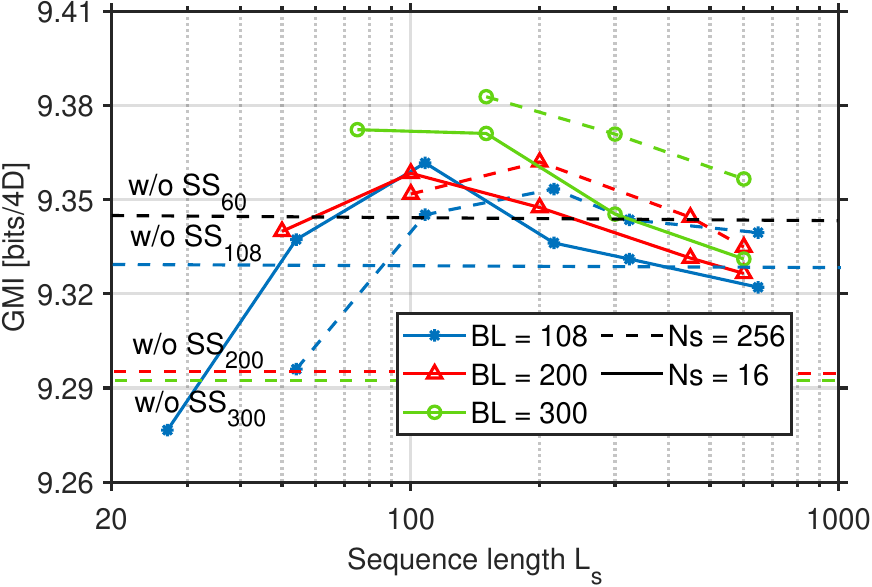}
    \caption{GMI versus selected sequence length. We show the optimal performance versus different sequence lengths, for various block lengths, by changing the number of flipping bits and the number of cascaded DMs. Same system configuration as in Fig.\ref{fig:SNR_elec_GMI_vs_complexity}. D-SS demonstrates superior performance over different sequence lengths.}
    \label{fig:Sequence_length_vs_BL}
\end{figure}

For a thorough analysis of the performance of D-SS, we vary the sequence length under identical system configuration and report the performance in Fig.~\ref{fig:Sequence_length_vs_BL}. We explore D-SS configurations with three different block lengths and vary the number of cascaded DMs, resulting in multiple sequence lengths. For each sequence length, we report the optimal performance obtained by adjusting the number of flipping bits and by performing the selection among $16$ (full lines) or $256$ (dashed lines) candidate sequences. For ease of comparison, the performance of ESS with four block lengths is also depicted on the graph using horizontal dashed lines. When the block length is set to $300$, D-SS shows the best performance due to lower rate loss and comparable nonlinear behavior with respect to shorter block lengths. $256$ selections tend to perform better for longer sequences because longer sequences necessitate cascading more DMs in series, hence providing more variability in the candidates for selection. With $256$ selections, D-SS can more effectively explore and utilize these possibilities. D-SS significantly outperforms ESS across almost the entire range of tested sequence lengths for block lengths of $200$ and $300$. D-SS with optimal sequence-length range ($100$ to $200$ 4D symbols) also surpasses the performance of ESS with block lengths of $60$ and $108$. The optimal sequence length for all four block lengths falls within the range of $100$ to $200$ 4D symbols. We explain the observed behavior by the fact that overly long sequences capture more channel memory, yet they lead to coarser selections (at a fixed number of selections to limit the complexity). On the other hand, while very short sequences offer finer selections, they only capture a small fraction of the channel memory. Consequently, for the specific transmission scenario discussed in this article, the balance point between channel memory and selection precision lies within the range of $100$ to $200$ 4D symbols.

\begin{figure}[t]
    \centering
    \includegraphics[width=1\linewidth]
    {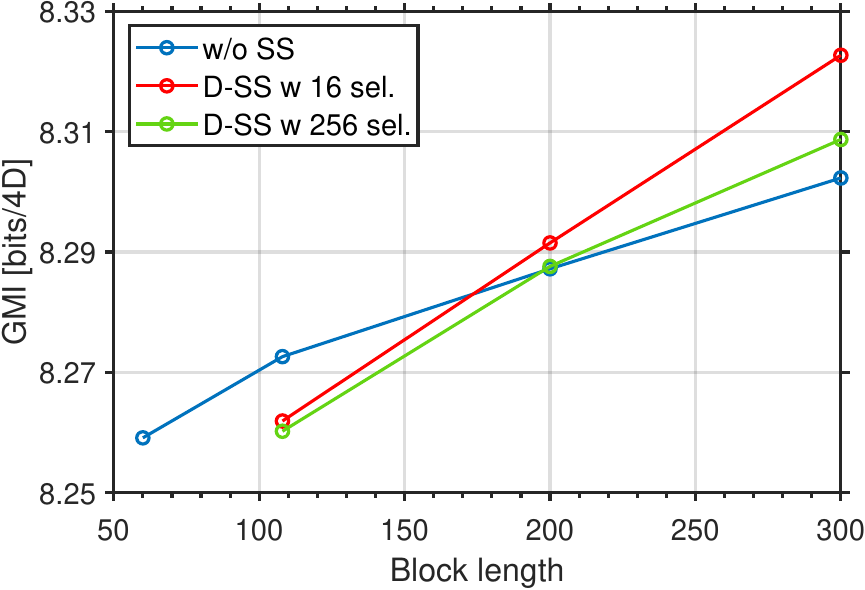}
    \caption{GMI versus DM block length. We show the optimal point versus block length by changing the number of flipping bits and the number of cascaded DMs. The same system configuration as in Fig.\ref{fig:SNR_elec_GMI_vs_complexity}, however with $5$ WDM channels. D-SS demonstrates superior performance in WDM transmission scenario.}
    \label{fig:performance_vs_BL_WDM}
\end{figure}

\subsection{WDM Transmission Performance}

Finally, we maintain the same system configuration as before but extend it to a $5$-channel WDM transmission. The combined net transmission rate for all five channels is $4.4$~Tbits/s. In our earlier investigation, we observed poor performance for D-SS with a block length of $60$. Consequently, we assess the performance starting with a block length of $108$. In Fig.~\ref{fig:performance_vs_BL_WDM}, we see that 
the nonlinear gain associated with short block lengths in ESS without SS diminishes because the CPR compensates for the increased inter-channel nonlinearity. There is an observed increase in GMI with increasing block length due to the reduced rate loss, yielding an optimal block length of $300$ for ESS. Conversely, D-SS better maintains nonlinear gains as the block length increases and its rate loss continues to decrease. This results in D-SS achieving the maximum GMI at a block length of $300$. Particularly, with $16$~selections, D-SS with a block length of $300$ outperforms ESS by $0.02$~bits/4D-symbol, and it achieves a $0.06$~bits/4D-symbol improvement compared to ESS with a block length of $60$. Lastly, it is important to note that $256$~selections, which require more flipping bits, lead to higher rate loss.  The additional nonlinear gain does not fully compensate for the performance degradation caused by the increased rate loss, resulting in a lower GMI than with $16$~selections. In a WDM configuration, the gains are lower than the ones observed in a single-channel system as a high number of cross-phase modulation components arise and our selection metric does not account for the energy dispersion of the neighboring channels. Finally, since these sequences are first selected for a single digital band and then used for all bands, the sequences selected by SSFM-SS are the same as D-SS and exhibit the same performance.

\section{Discussion and Conclusion}
\label{Conclusion}

We have proposed a novel sign-dependent metric: D-EDI, which can serve as a prediction metric for the generation of NLI because it showed a negative correlation with the measured performance of long-distance transmission systems. By applying D-EDI (resp. EDI) in sequence selection, we defined the D-SS (resp. E-SS) scheme. These two schemes achieved higher gains with ESS used for distribution matching than state-of-the-art ESS schemes over short transmission distances. Noticeably, D-SS showed gains even for longer transmission distances and higher baud rates over single-channel and WDM systems even when CPR is used to reduce the impact of nonlinear phase rotations, achieving the same performance as sequence selection based on a full SSFM simulation~\cite{civelli2023sequence}. Moreover, D-SS retained a robust performance at moderate complexity (lower number of digital dispersion operations to compute the metric) and outperformed conventional ESS across different block lengths. These results pave the way for further optimizations of shaping schemes for long-distance transmissions and confirm the necessity of using sign-dependent and channel-aware metrics for effectively evaluating nonlinear distortions.

Nevertheless, these findings tell us that there is still room for improvement. For DMB-based transmissions in which the transmission rate per subcarrier is lower than single-carrier transmissions, the impact of channel memory is limited to a few symbols per span or even less than 1 symbol per span, hence only a few dozen symbols for all spans. Thus, the impact of channel memory can be accounted for by concatenating several DMs to minimize the mutual influence of adjacent selected sequences. As the transmission rates continue to increase, for both multi-carrier and single-carrier transmissions, the channel memory can spread over hundreds or thousands of symbols. At the same time, increasing the length of candidate sequences also increases the complexity of the scheme. Hence, modifications and optimizations of D-EDI to achieve excellent performance predictions under high-rate long-distance transmission while maintaining a reduced complexity is still an open question. Other improvements of the metric may address different dispersion management schemes or take into account inter-channel nonlinear effects. In this work, we only considered long-haul dispersion-unmanaged systems where the NLI contributions from different spans can be considered independent and D-EDI gives a good performance prediction by summing up EDIs of the sequence at the start of each span.

Ultimately, an extended goal involves conducting an in-depth analysis of the selected ``good" sequences. By comprehensively understanding their properties, we aim to design simpler sequence generation solutions.

\section*{Acknowledgments}
This work was funded by Huawei Technologies France. We thank Yann Frignac for fruitful discussions.


 


\bibliography{References}{}
\bibliographystyle{IEEEtran}

\end{document}